# TESTING SIGNIFICANCE OF FEATURES BY LASSOED PRINCIPAL COMPONENTS

By Daniela M. Witten[1] and Robert Tibshirani[2]

*Stanford University*

We consider the problem of testing the significance of features in high-dimensional settings. In particular, we test for differentially-expressed genes in a microarray experiment. We wish to identify genes that are associated with some type of outcome, such as survival time or cancer type. We propose a new procedure, called Lassoed Principal Components (LPC), that builds upon existing methods and can provide a sizable improvement. For instance, in the case of two-class data, a standard (albeit simple) approach might be to compute a two-sample $t$-statistic for each gene. The LPC method involves projecting these conventional gene scores onto the eigenvectors of the gene expression data covariance matrix and then applying an $L_1$ penalty in order to de-noise the resulting projections. We present a theoretical framework under which LPC is the logical choice for identifying significant genes, and we show that LPC can provide a marked reduction in false discovery rates over the conventional methods on both real and simulated data. Moreover, this flexible procedure can be applied to a variety of types of data and can be used to improve many existing methods for the identification of significant features.

**1. Introduction.** In recent years new experimental technologies within the field of biology have led to data sets in which the number of features $p$ greatly exceeds the number of observations $n$. Two such examples are gene expression data and data from genome-wide association studies. In the case of gene expression (or microarray) data, it is often of interest to identify genes that are differentially-expressed across conditions (for instance, some patients may have cancer and others may not), or that are associated with some type of outcome, such as survival time. Such genes might be used as

Received February 2008; revised May 2008.

[1]Supported by a National Defense Science and Engineering Graduate Fellowship.

[2]Supported in part by NSF Grant DMS-99-71405 and the National Institutes of Health Contract N01-HV-28183.

features in a model for prediction or classification of the outcome in a new patient, or they might be used as target genes for further experiments in order to better understand the biological processes that contribute to the outcome.

Over the years, a number of methods have been developed for the identification of differentially-expressed genes in a microarray experiment; for a review, see Cui and Churchill (2003) or Allison et al (2006). Usually, the association between a given gene and the outcome is measured using a statistic that is a function of that gene only. Genes for which the statistic exceeds a given value are considered to be differentially-expressed. Many methods combine information, or borrow strength, across genes in order to make a more informed assessment of the significance of a given gene. In the case of a two-class outcome, such methods include the shrunken variance estimates of Cui et al. (2005), the empirical Bayes approach of Lonnstedt and Speed (2002), the `limma` procedure of Smyth (2004) and the optimal discovery procedure (ODP) of Storey, Dai and Leek (2007). We elaborate on the latter two procedures, as we will compare them to our method throughout the paper in the case of a two-class outcome. Limma assesses differential expression between conditions by forming a moderated $t$-statistic in which posterior residual standard deviations are used instead of the usual standard deviation. The ODP approach is quite different; it involves a generalization of the likelihood ratio statistic such that an individual gene's significance is computed as a function of all of the genes in the data set. In the case of a survival outcome, a standard method to assess a gene's significance (and the one to which we will compare our proposed method in this paper) is the Cox score; see, for example, Beer et al. (2002) and Bair and Tibshirani (2004).

We propose a new method, called Lassoed Principal Components (LPC), for the identification of differentially-expressed genes. The LPC method is as follows. First, we compute scores for each gene using an existing method, such as those mentioned above. These scores are then regressed onto the eigenarrays of the data [Alter, Brown and Botstein (2000)]—principal components of length equal to the number of genes—with an $L_1$ constraint. The resulting fitted values are the LPC scores. In this paper we demonstrate that LPC scores can result in more accurate gene rankings than the conventional methods, and we present theoretical justifications for the use of the LPC method.

Our method has two main advantages over existing methods:

1. LPC borrows strength across genes in an explicit manner. This benefit is rooted in the biological context of the data. In biological systems genes that are involved in the same biological process, pathway, or network may be co-regulated; if so, such genes may exhibit similar patterns of expression. One would not expect a single gene to be associated with the outcome, since, in practice, many genes work together to effect a particular phenotype. LPC effectively down-weights individual genes that are associated with the outcome but that do not share an expression pattern with a larger group of genes, and instead favors large groups of genes that appear to be differentially-expressed. By implicitly using prior knowledge about what types of genes one expects to be differentially-expressed, LPC achieves improved power over existing methods in many examples.

2. LPC can be applied on top of any existing method (regardless of outcome type) in order to obtain potentially more accurate measures of differential expression. For instance, in the case of a two-class outcome, many methods to detect differentially-expressed genes exist, including SAM [Tusher, Tibshirani and Chu (2001)], limma [Smyth (2004)] and ODP [Storey, Dai and Leek (2007)], as mentioned earlier. By projecting any of these methods' gene scores onto the eigenarrays of the data with an $L_1$ constraint, we harness the power of these methods while incorporating the biological context.

The idea behind this method is that a gene that resembles the outcome is more likely to be significant if it is one of many genes with similar expression patterns than if it resembles no other genes. From a biological standpoint, this is due to the hypothesis that a gene that truly is associated with an outcome (such as cancer) will be involved in a biological pathway related to the outcome that involves many genes. Other genes in the pathway may exhibit similar expression patterns due to co-regulation. From a statistical standpoint, it is due to the fact that while variance in the genes' expression levels may occasionally cause an individual nonsignificant gene to be correlated with the outcome by chance, it is statistically quite unlikely that a great number of genes will all be correlated with the outcome and with each other due solely to random noise. By regressing the conventional gene scores onto the eigenarrays of the gene expression data and using the fitted values to rank genes, we essentially only rank highly the genes that have moderate to high gene scores and have large loadings in an eigenarray that is correlated with the vector of gene scores. Thus, individual genes with expression patterns that do not resemble those of other genes in the data set are not given high rankings by our method. Genes with moderate scores that resemble a large block of genes with high scores are given high LPC scores; they borrow strength from genes with similar expression profiles.

The LPC method bears similarities to the surrogate variable analysis (SVA) method of Leek and Storey (2007). SVA attempts to adjust for expression heterogeneity among samples, whereas we try to exploit heterogeneity in order to obtain more accurate gene scores. The effects of the two methods are quite different, and the methods can be used together, as is shown in Appendix D.2.

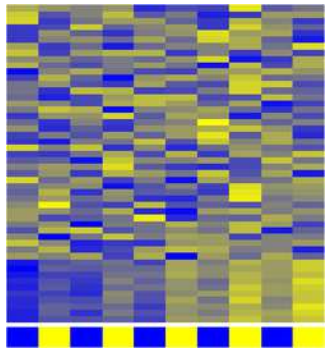

Fig. 1. *Heatmap of simulated data set; blue signifies low and yellow signifies high gene expression. There are 50 genes (rows) and 10 patients (columns); 10 of the genes (shown on the bottom of the heatmap) are associated with the outcome. Patients are ordered from oldest to youngest. The binary outcome for each patient is shown below the heatmap. The alternating pattern of the outcome can be seen in the significant genes in the heatmap, but the predominant pattern seen in these genes is associated with age.*

A simulated two-class example helps to illustrate how LPC can outperform standard approaches. Suppose that expression profiles come from either cancer or normal tissue, and that genes over-expressed in cancer also happen to be under-expressed in older individuals (Figure 1, see Supplementary Materials for R language code for this simulation [Witten and Tibshirani (2008)]). The expression of these genes is a function of both patient age and tissue type. If it is known by the data analyst that age affects gene expression, then age can be used as a covariate in determining whether a gene is differentially-expressed. However, in practice, factors that affect gene expression are often unknown or unmeasured, and so are not included as covariates. In this case, a two-sample $t$-statistic will have limited success in identifying the genes associated with cancer type, because the age effect masks some of the correlation between cancer type and gene expression. On the other hand, applying the LPC method to the two-sample $t$-statistics results in high scores for the differentially-expressed genes, because these genes will have high loadings on the eigenarray that is most correlated with the cancer type. The resulting gene scores can be seen in Figure 2.

LPC is not restricted to the two-class problem, and findings in the context of survival outcomes indicate its potential promise. Figure 3 shows the estimated false discovery rates in detecting genes in lymphoma patients that are associated with altered survival. LPC clearly outperforms a standard gene-specific analysis based on Cox scores (see Section 3.3).

The paper is organized as follows. In Section 2 we present the details of the LPC method, as well as some theoretical results that justify its use. Then, in Section 3 we demonstrate by example that LPC outperforms the conventional gene scores in simulations and on published microarray data sets for two-class and survival outcomes. We also present a method for false discovery rate estimation for LPC, which is given in greater detail in Appendix C. Section 4 contains the Discussion.

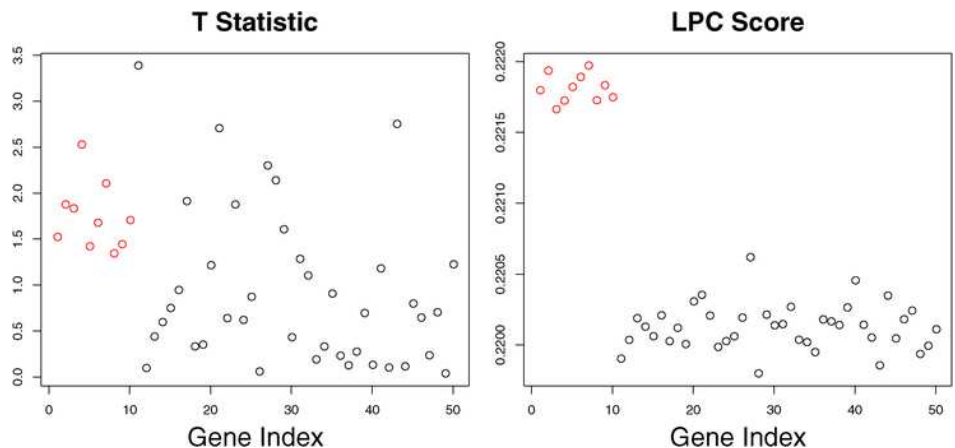

Fig. 2. *Genes that truly are differentially-expressed are shown in red. The two-sample t-statistic does not assign very high scores to the differentially-expressed genes, because the confounding effect of age reduces the association between these genes and the outcome. LPC (based on the two-sample t-statistic) assigns high scores to all of the significant genes.*

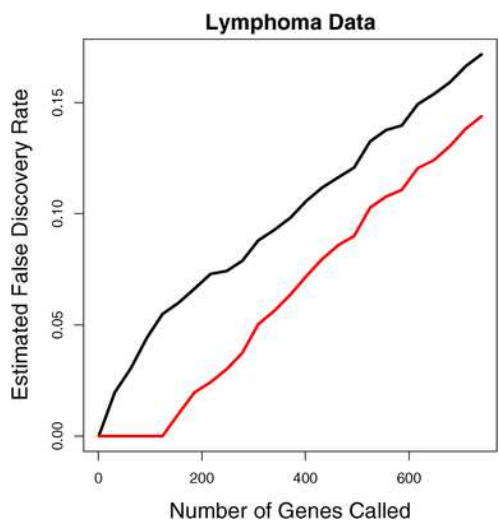

Fig. 3. *Estimated false discovery rates are shown for the lymphoma data set [Rosenwald et al. (2002)], which consists of 7,399 gene expression measurements on 240 patients. The false discovery rate of Cox scores is shown in black, and that of LPC is shown in red.*

TABLE 1
*Simple scores for various outcomes*

| Outcome type | Gene score |
|---|---|
| Quantitative | Standardized regression coefficient for regression of $\mathbf{y}$ onto $\mathbf{X}_j$ |
| Survival | Score statistic for univariate Cox proportional hazards model; see, for example, Beer et al. (2002) and Bair and Tibshirani (2004) |
| Two-class | Two-sample $t$-statistic |
| Multiple-class | F-statistic for one-way ANOVA |

## 2. The LPC method.

2.1. *Description.* Let $\mathbf{X}$ denote an $n \times p$ matrix of log transformed gene expression levels, where $n$ is the number of observations and $p$ is the number of genes, and $n \ll p$. Assume that the columns of $\mathbf{X}$ (the genes) have been centered so that they have mean zero across all of the observations. Let $\mathbf{X}_j$ denote the vector of expression levels for gene $j$. Let $\mathbf{y}$ denote a vector of length $n$ containing the outcomes for each observation. For instance, if this is a two-class problem, then $\mathbf{y}$ will be a binary vector.

The LPC method involves using existing gene scores to develop LPC scores that aim to provide a more accurate ranking of genes in terms of differential-expression. In principle, a wide variety of gene scores could be used; however, in the simplest version of LPC one would use one of the methods in Table 1, depending on the outcome variable. In the examples analyzed, a small constant is added to the denominators of the gene scores in Table 1 in order to avoid large ratios resulting from small estimated standard deviations; see, for example, Tusher, Tibshirani and Chu (2001). We will refer to the statistics in Table 1 as $T$. For ease of notation, unless we specify otherwise, the LPC scores discussed in this paper are formed by applying the LPC method to these $T$ scores. $T_j$ will refer to the gene score for gene $j$. The LPC method is as follows for the quantitative, survival and two-class cases:

**LPC Algorithm:**

1. Compute $\mathbf{T}$, the vector of length $p$ with components $T_j$, the score for gene $j$, for $j \in 1, \ldots, p$.
2. Compute $\mathbf{v}_1, \ldots, \mathbf{v}_n$, the eigenarrays of $\mathbf{X}$, where the $\mathbf{v}_i$'s are the columns of the matrix $\mathbf{V}$ in the singular value decomposition (SVD) of $\mathbf{X}$, $\mathbf{X} = \mathbf{U}\mathbf{D}\mathbf{V}^T$.
3. For some value of the parameter $\lambda$, fit the model $\mathbf{T} = \boldsymbol{\beta}_0 + \sum_{i=1}^n \beta_i \mathbf{v}_i$, where the vector $\boldsymbol{\beta}$ with components $\beta_i$ is chosen to minimize the quantity $(\mathbf{T} - \boldsymbol{\beta}_0 - \mathbf{V}\boldsymbol{\beta})^T(\mathbf{T} - \boldsymbol{\beta}_0 - \mathbf{V}\boldsymbol{\beta}) + \lambda \sum_{i=1}^n |\beta_i|$. This is multiple linear regression with an $L_1$ constraint, also known as the "lasso" [Tibshirani (1996)].

4. Let $\hat{\mathbf{T}}$ denote the fitted values obtained by the above model. The LPC score for gene $j$ is $\hat{T}_j$.

In the case of a multiple-class response, the procedure is slightly different, and is presented in Appendix B.

In Step 3 of the LPC algorithm, fitting a linear model with an $L_1$ constraint is very fast, because we are regressing the scores $\mathbf{T}$ on the columns of $\mathbf{V}$, which are orthogonal. We use the following soft thresholding approach in order to obtain the lasso coefficients:

**Soft-Thresholding Algorithm:**

1. Compute $\hat{\boldsymbol{\beta}}$, the vector of coefficients obtained by regressing $\mathbf{T}$ on the eigenarrays of $\mathbf{X}$ using ordinary multiple least squares; that is, $\hat{\boldsymbol{\beta}} = \arg\min_{\boldsymbol{\beta}} (\mathbf{T} - \boldsymbol{\beta}_0 - \sum_{i=1}^{n} \mathbf{v}_i \beta_i)^T (\mathbf{T} - \boldsymbol{\beta}_0 - \sum_{i=1}^{n} \mathbf{v}_i \beta_i)$.
2. Let $\tilde{\beta}_i = \text{sign}(\hat{\beta}_i)(|\hat{\beta}_i| - \frac{\lambda}{2})_+, \forall i \in 1, \ldots, n$.
3. Compute $\hat{\mathbf{T}} = \hat{\boldsymbol{\beta}}_0 + \sum_{i=1}^{n} \mathbf{v}_i \tilde{\beta}_i$; these are the LPC scores.

The LPC algorithm involves a shrinkage parameter, $\lambda$, which determines the amount of regularization performed in the $L_1$-constrained regression. An automated method for the selection of the value for this parameter is presented in Appendix A.

Returning to the example from the Introduction, the value of the shrinkage parameter $\lambda$ chosen by our automated method was 5.5. This resulted in $\hat{\beta}_1$ nonzero and $\hat{\beta}_i = 0$ for $i \in 2, \ldots, n$. The first eigenarray is associated with the response. In this example, LPC's success stems from the fact that the $L_1$ constraint resulted in a nonzero coefficient only for the correct eigenarray.

In the case of a quantitative response, the $T$ scores take the form $\frac{\mathbf{X}_j^T \mathbf{y}}{\sigma \sqrt{(\mathbf{X}^T \mathbf{X})_{jj}}}$ for gene $j$. Suppose that the genes have been scaled appropriately so that the $T$ scores are simply $\mathbf{X}^T \mathbf{y}$. From the LPC Algorithm and the Soft-Thresholding Algorithm, the LPC scores are given by the formula $\hat{\mathbf{T}} = \hat{\boldsymbol{\beta}}_0 + \sum_{i=1}^{n} \mathbf{v}_i \tilde{\beta}_i$, where the columns of $\mathbf{V}$ are linear combinations of the rows of $\mathbf{X}$. Therefore, if $\lambda = 0$ (i.e., in the absence of an $L_1$ constraint), the LPC scores equal the $T$ scores exactly. This means that $T$ is a special case of LPC. This leads us to hope that if, on a given data set, $T$ outperforms LPC with nonzero $\lambda$, our adaptive method of choosing $\lambda$ will set $\lambda$ to zero. If this is the case, then we will always end up using the approach that works best on a particular data set. A similar result holds for the case of a two-class response. Note that, in practice, however, one usually does not scale the genes as described here.

2.2. *Motivating LPC via an underlying latent variable model.* Consider a scenario in which a subset of genes is associated with the outcome because some underlying process, or "latent variable," affects both the expression of the genes and the outcome measurements. In Appendix D.1, it is shown that in such a situation, under suitable assumptions, LPC scores will have lower variance than $T$ scores. This justifies the use of LPC in situations where a latent variable model could reasonably describe the data set of interest.

2.3. *Relationship with the eigengene space.* In microarray data analysis the principal components of the columns of $\mathbf{X}$ are referred to as the eigengenes, and the principal components of the rows of $\mathbf{X}$ are referred to as the eigenarrays. We are interested in identifying significant genes; therefore, it may seem peculiar that our proposed method works in the space of eigenarrays rather than in the space of eigengenes. For instance, Bair and Tibshirani (2004) and Bair et al. (2006) perform supervised principal components analysis in the eigengene space. We show here that, in the simple case of a quantitative outcome, working in the eigenarray space is quite similar to working in the eigengene space, but has a distinct advantage.

Assume that the columns of $\mathbf{X}$ are centered and that the quantitative outcome $\mathbf{y}$ is centered. Let $\mathbf{X} = \mathbf{U}\mathbf{D}\mathbf{V}^T$ denote the singular value decomposition for $\mathbf{X}$. Let $\mathbf{T}$ denote the vector of $T$ statistics. The LPC method fits the linear model with $L_1$ constraint on the coefficients

$$(2.1) \qquad \mathbf{T} = \mathbf{V}\boldsymbol{\beta} + \boldsymbol{\varepsilon} = \sum_{i=1}^{n} \mathbf{v}_i \beta_i + \boldsymbol{\varepsilon},$$

where $\mathrm{E}(\boldsymbol{\varepsilon}) = 0$ and $\mathbf{v}_i$ is the $i$th right singular vector of the gene expression data; in other words, it is the $i$th eigenarray of the original data $\mathbf{X}$.

Now, suppose that instead of using the conventional $T$ scores (which, in this case, would be the standardized regression coefficients), we take the inner product of each gene $\mathbf{X}_j$ with the vector $\mathbf{y}$. (If the genes of $\mathbf{X}$ were scaled appropriately before computing the usual $T$ scores, then the usual $T$ scores would be equivalent to these simplified scores.) The above equation gives

$$(2.2) \qquad \begin{aligned} \mathbf{T} &= \mathbf{X}^T \mathbf{y} \\ &= \mathbf{V}\boldsymbol{\beta} + \boldsymbol{\varepsilon} \\ &= \mathbf{X}^T \mathbf{U}\mathbf{D}^{-1}\boldsymbol{\beta} + \boldsymbol{\varepsilon} \\ &= \mathbf{X}^T \mathbf{U}\boldsymbol{\theta} + \boldsymbol{\varepsilon}, \end{aligned}$$

letting $\boldsymbol{\theta}$ be a vector of length $n$ with components $\theta_i = \frac{\beta_i}{d_i}$, where $d_i$ is the $i$th diagonal element of $\mathbf{D}$.

Assuming that $\lambda = 0$ (so we are simply performing multiple least squares regression), then

$$\boldsymbol{\theta} = \arg\min \|\mathbf{X}^T \mathbf{y} - \mathbf{X}^T \mathbf{U}\boldsymbol{\theta}\|^2$$

$$(2.3) \qquad = \arg\min \|\mathbf{X}^T(\mathbf{y} - \mathbf{U}\boldsymbol{\theta})\|^2$$
$$= \arg\min \|\mathbf{D}\mathbf{U}^T(\mathbf{y} - \mathbf{U}\boldsymbol{\theta})\|^2.$$

Note that if $\mathbf{X}\mathbf{X}^T = \mathbf{U}\mathbf{D}^2\mathbf{U}^T = \mathbf{I}$, then

$$(2.4) \qquad \boldsymbol{\theta} = \arg\min \|\mathbf{y} - \mathbf{U}\boldsymbol{\theta}\|^2.$$

Now, suppose that we instead regress $\mathbf{y}$ on the columns of $\mathbf{U}$:

$$(2.5) \qquad \mathbf{y} = \sum_{i=1}^{n} \mathbf{u}_i \theta'_i + \varepsilon'$$
$$= \mathbf{U}\boldsymbol{\theta}'.$$

Here, if $\lambda = 0$, then $\boldsymbol{\theta}'$ is again given by (2.4).

So, if $\lambda = 0$, then regressing the scores on the eigenarrays is quite similar to regressing the outcome on the eigengenes. In fact, we have shown that if $\mathbf{X}\mathbf{X}^T = \mathbf{I}$, then the resulting coefficients are identical up to a scaling by the matrix $\mathbf{D}$ of the SVD.

In general, when $\lambda = 0$, the solution to the LPC least squares problem is the $\boldsymbol{\theta}$ that minimizes

$$(2.6) \qquad \|\mathbf{X}^T(\mathbf{y} - \mathbf{U}\boldsymbol{\theta})\|^2 = \|\tilde{\mathbf{y}} - \tilde{\mathbf{U}}\boldsymbol{\theta}\|^2,$$

where $\tilde{\mathbf{y}} = \mathbf{D}\mathbf{U}^T\mathbf{y}$, $\tilde{\mathbf{U}} = \mathbf{D}\mathbf{U}^T\mathbf{U}$. Therefore, in the case of a quantitative response with simplified $T$ scores of the form $\mathbf{X}^T\mathbf{y}$, regressing the scores on the eigenarrays is equivalent to regressing the outcomes on the eigengenes after rotating the outcome and the eigengenes.

When the outcome is quantitative, working in the eigengene space is a reasonable alternative to working in the eigenarray space. However, in the case of a nonquantitative outcome, working in the eigengene space does not always generalize. For instance, suppose that the outcome is survival time. Then, for each observation, the response is a time (e.g., number of months that the patient survived) and a binary variable (whether or not the patient was censored). It is not clear how to regress this pair of numbers onto the eigengenes. However, using LPC, we first compute the Cox scores; it is then a simple matter to regress the Cox scores onto the eigenarrays. Therefore, working in the eigenarray space has a distinct advantage in that it is applicable to a wider range of outcome types.

## 3. Performance of LPC.

3.1. *Simulated data.* LPC outperforms existing methods in a variety of simulations, for quantitative, survival, two-class, and multiple-class responses. We present results on three such simulations here. Each simulation involves 1000 genes and 40 observations. The first 50 genes are associated with the outcome for each observation, and will be referred to as the significant genes. Simulation 1 represents the simplest case; the only structure present in the data is due to the differentially-expressed genes, which closely resemble the outcome. Simulation 2 is more complicated: in addition to the genes that resemble the outcome, there are three blocks of 100 genes that display very strong expression patterns that are orthogonal to the outcome. In Simulation 3, the 50 genes that are associated with the outcome are split into two sets of 25 genes; the signals present in each set are orthogonal to each other, and the response is obtained by summing the two orthogonal signals. Detailed descriptions of each simulation can be found in Appendix E, and heatmaps and R language code can be found in the Supplementary Materials [Witten and Tibshirani (2008)].

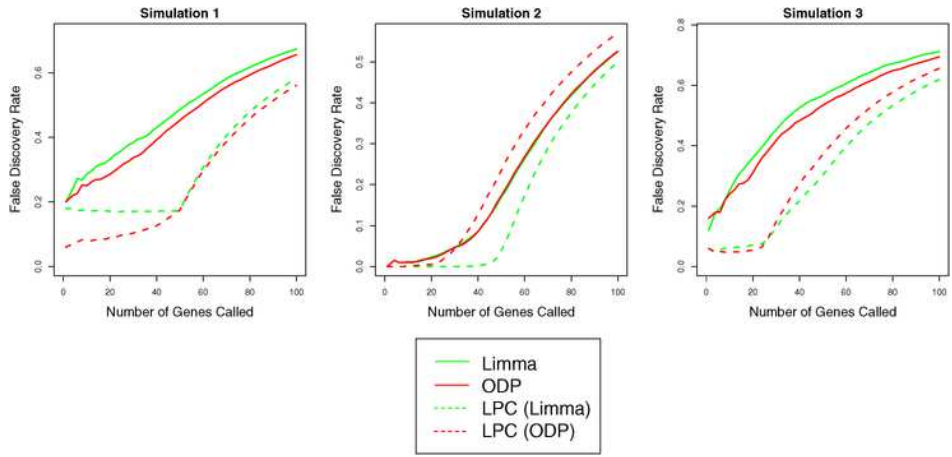

FIG. 4. *False discovery rates are shown for Simulations 1, 2 and 3 with a two-class response. In each simulation, 50 out of 1000 genes are significant. ODP, LPC(ODP), Limma and LPC(Limma) are compared. In almost all cases, the LPC methods result in lower FDRs than the existing methods. T and LPC(T), though not plotted, closely resemble the Limma and LPC(Limma) curves for all three figures.*

The performances of ODP [Storey, Dai and Leek (2007)] and limma [Smyth (2004)] are compared to the performances of the LPC method applied to ODP and limma for each simulation (with a two-class outcome) in Figure 4. It is clear that the statistics are improved by applying LPC. Though not shown in the figure, $T$ is also improved by applying the LPC method. Similar results hold for other outcome types.

It is worth noting that the $L_1$ constraint in the LPC algorithm is an important part of the reason that LPC outperforms the statistics to which it is applied. A reviewer inquired whether one would expect to see the same gains if one simply projected the initial statistics onto, say, the first $k < n$ eigenarrays. Consider Simulation 2 as described above (with a two-class response), but assume that now there are seven, rather than three, blocks of 100 noise genes that are orthogonal to the response. We can compare the LPC scores to the scores obtained by regressing the $T$ scores onto the first five eigenarrays. As expected, Figure 5 shows that this latter method results in very poor performance, since the blocks of noise genes dominate the first five principal components. On the other hand, LPC does well because the $L_1$ constraint leads to sparsity in the regression coefficients, and so only the eigenarray or eigenarrays that are related to the response are included in the model.

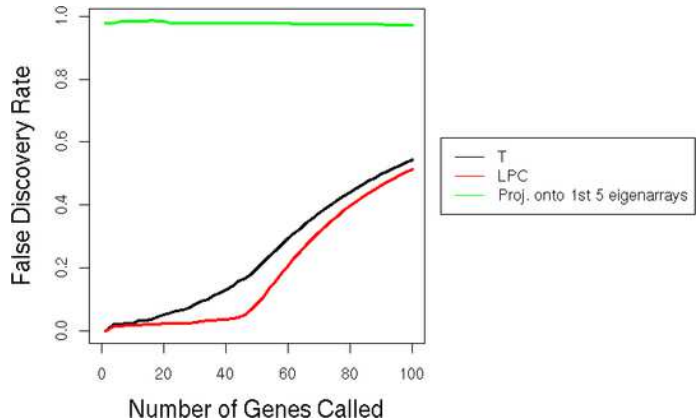

Fig. 5. *False discovery rates are shown for a simulation with seven blocks of noise genes that are orthogonal to the two-class outcome. LPC performs much better than simply regressing T onto the first five eigenarrays.*

3.2. *Predictive advantage.* An objection to the use of LPC might be that despite its performance on simulated data and its theoretical properties, it simply does not rank genes using the metric that is truly of interest. For instance, in the case of two-class data, a biologist might truly care about finding genes that have different mean expression in the two classes. Such a researcher might claim that $T$ succinctly quantifies the concept of interest, whereas LPC finds genes that satisfy the rather abstract and perhaps irrelevant requirement of having $T$ scores which, when projected onto the high-variance subspace of the gene expression data, yield large values. Here we show that in many cases, even if the conventional $T$ scores are the quantities of interest, LPC should be used instead of $T$.

Let $\mathbf{X}$ and $\mathbf{X}^*$ denote independent training and test sets for the same set of genes. $\mathbf{T}$ and $\mathbf{L}$ are the vectors of conventional $T$ scores and LPC scores on the training set, and $\mathbf{T}^*$ is the vector of conventional $T$ scores on the test

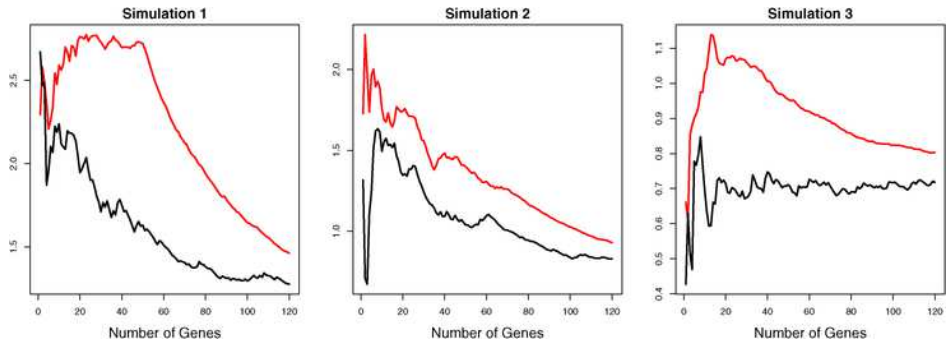

FIG. 6. $\mathrm{E}(|T_j^*|\,|\,|L_j| > c_2(\alpha))$ *(red) and* $\mathrm{E}(|T_j^*|\,|\,|T_j| > c_1(\alpha))$ *(black) are shown for Simulations 1, 2 and 3 with a quantitative outcome. The difference between the two quantities gives the predictive advantage. The predictive advantage is positive in all cases. Similar results hold for other outcome types.*

set. Then, we define *predictive advantage* as

$$(3.1) \qquad \mathrm{E}(|T_j^*|\,|\,|L_j| > c_2(\alpha)) - \mathrm{E}(|T_j^*|\,|\,|T_j| > c_1(\alpha)),$$

where $c_1(\alpha)$ and $c_2(\alpha)$ are the $\alpha$ quantiles of $|T_j|$ and $|L_j|$, respectively. A positive predictive advantage for LPC essentially means that even if $T$ is the quantity of interest, then LPC should be used instead, since LPC will pick out genes with higher $T$ scores on an independent test set. If LPC has a positive predictive advantage on a given data set, then there is no question that LPC is superior to $T$ on that data set.

On data that we examined, the predictive advantage is often positive (Figures 6 and 7). The simulated data for Figure 6 has a quantitative outcome; for each simulation, the predictive advantage is positive. Two of the data sets used for Figure 7 have survival outcomes; they are the lymphoma data set from Rosenwald et al. (2002) and the kidney cancer data set from Zhao et al. (2006). The third data set is the two-class colon cancer data set of Alon et al. (1999). The predictive advantage of LPC is positive for the survival data sets, and is positive after more than the first few genes have been selected for the two-class data set.

The predictive advantage provides a "quick and dirty" approach to verifying that LPC is uniformly better at identifying significant genes on a given data set. We recommend the use of LPC instead of competing methods whenever its predictive advantage relative to the competing methods is positive. However, the predictive advantage of LPC was not positive for all of the data sets that we considered.

3.3. *Estimated false discovery rates for LPC.* Estimation of false discovery rates for LPC is surprisingly difficult. Due to the fact that the LPC statistic for a given gene is a function of all of the genes present in the data set, the standard method of estimating false discovery rates by permutations cannot be applied. More on this topic, as well as the method that we developed to estimate false discovery rates for LPC and other functionally dependent statistics using predictive advantage, can be found in Appendix C.

We apply our method of estimating false discovery rates for LPC to the kidney cancer and lymphoma survival data sets of Zhao et al. (2006) and Rosenwald et al. (2002), as well as to the two-class colon data set of Alon et al. (1999). For the survival data sets, we compare LPC to $T$, and for the two-class data set, we compare $T$, limma, ODP and LPC. Figures 3 and 8 show that the estimated FDR of LPC is less than those of the other methods, with the exception of the first 15 or so genes for the colon cancer data set. The erratic performance of LPC for these first genes is a direct consequence of the fact that the predictive advantage of LPC is negative for these first genes in this data set (see Figure 7).

As discussed in Appendix C, our method of estimating FDRs for LPC involves computing FDRs for simpler scores that can be estimated through permutations, and then estimating the difference in FDR between LPC and those simpler scores. Figure 9 displays the mean estimated difference between the FDR of $T$ and the FDR of LPC, as well as standard errors for this estimate, for the colon, kidney, and lymphoma data sets. The figure was obtained by repeatedly sampling 90% of the observations, without replace-ment; for each resulting sample, the difference in FDR between $T$ and LPC was computed. The figure shows that for these three data sets, the estimated FDR of $T$ is significantly higher than that of LPC (with the exception of the first few genes in the colon data set).

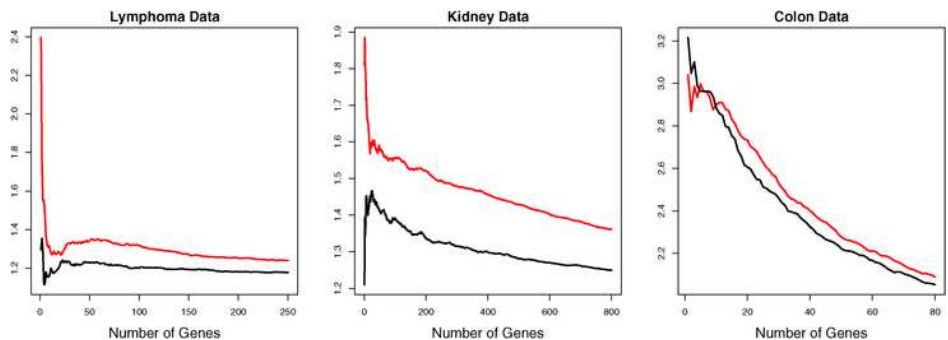

FIG. 7. $\mathrm{E}(|T_j^*| \mid |L_j| > c_2(\alpha))$ (red) and $\mathrm{E}(|T_j^*| \mid |T_j| > c_1(\alpha))$ (black) are shown for two survival data sets: the kidney data [Zhao et al. (2006)] and the lymphoma data [Rosenwald et al. (2002)]. In both cases, the predictive advantage of LPC is positive. The two-class colon data [Alon et al. (1999)] is also shown; in this case, the predictive advantage is positive when more than around 15 genes are considered. The kidney data set involves survival times and expression measurements on 14,814 genes for 177 patients. As mentioned earlier, the lymphoma data set involves survival times and expression measurements on 7,399 genes for 240 patients. The colon data consists of cancer status and gene expression measurements on 2,000 genes for 62 patients.

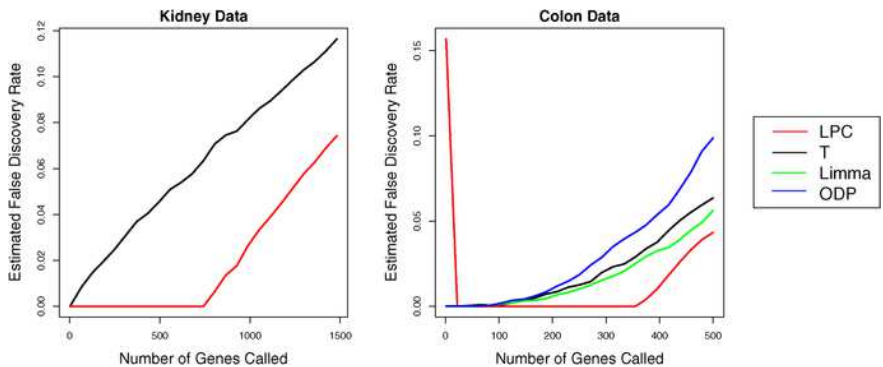

Fig. 8.    *Estimated FDR for two published microarray data sets: the Zhao et al. (2006) kidney cancer survival data set and the Alon et al. (1999) colon cancer two-class data set.*

3.4. *Another look at LPC versus $T$ for survival data.* First, we examine more closely the differences between the genes selected as significant by LPC and $T$ (in this case, Cox scores) in the kidney cancer data set of Zhao et al. (2006). When LPC is applied to this data set, several eigenarrays (out of 177) are assigned nonzero coefficients; two of these, eigenarrays 2 and 4, have large absolute coefficients. Figure 10 shows the loadings of the genes on these two eigenarrays. The top 1% of genes selected by LPC and the top 1% of genes selected by Cox scores are shown. Though there is some overlap between these two gene sets, it is clear from the figure that LPC selects genes with extreme loadings on these two eigenarrays, whereas the Cox score uses a different criterion.

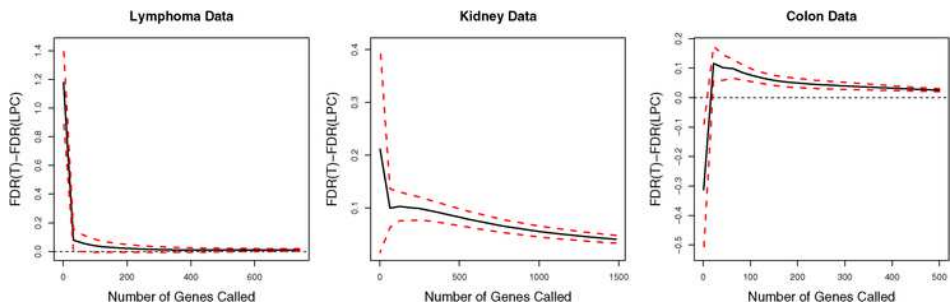

Fig. 9.    *The mean estimated difference in FDR between $T$ and LPC is shown in black, as a function of the number of genes called significant, for 20 data sets created by sampling 90% of the observations from the original data set without replacement. An estimate of one standard error above and below this mean is shown in red. A positive mean difference indicates that $T$ has higher FDR than LPC.*

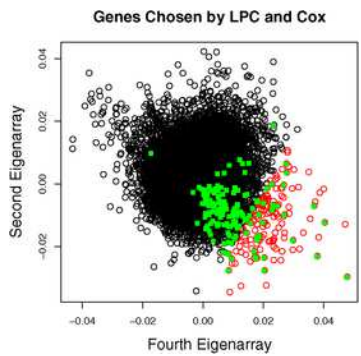

FIG. 10. *The loadings of the kidney cancer genes on the second and fourth eigenarrays are shown. The top 1% of genes selected by LPC are shown in red, and the top 1% of genes selected by Cox score are shown in green. The remaining genes are in black. There is some overlap between the genes with the highest Cox scores and the genes with the highest LPC scores.*

The genes with the highest LPC scores have high loadings on eigenarrays 2 and 4; this means that they have expression that is correlated with eigengenes 2 and 4. Figure 11 shows the top genes selected by LPC and not by Cox score, the top genes selected by Cox score and not by LPC, and the fourth eigengene. It is clear that the top genes selected by LPC and not by Cox score resemble this eigengene more closely than do the genes selected by Cox score and not by LPC.

We established earlier that in the kidney cancer and lymphoma data sets, LPC has a positive predictive advantage relative to Cox scores. We now examine a related concept, the ability of the top genes ranked by LPC and Cox scores to predict survival. We split the kidney and lymphoma data sets into a training set and a test set, and identified the 25 genes with highest Cox and LPC scores on the training set. We fit Cox proportional hazard models to the test data, using the top $i = 1, \ldots, 25$ genes ranked by each scoring method. We then computed the medians of the log rank statistics obtained from each of these models over 20 iterations. The results can be seen in Figure 12. Models that use the top genes ranked by LPC outperform models that use the top genes ranked by Cox scores when not too many genes are included in the model. However, when enough genes are included in the model, then for both data sets, genes with high training set Cox scores eventually lead to superior models. This is due to the fact that LPC assigns high scores to correlated sets of genes, and including additional correlated predictors in a model does not lead to much improvement. On the other hand, using Cox scores results in the addition of genes to the model that may be less correlated with each other, and so adding a greater number of such genes leads to greater improvement. This example demonstrates that while we have shown that LPC is more successful than competing methods at identifying significant genes, it is not ideally suited for model selection if a large model is desired.

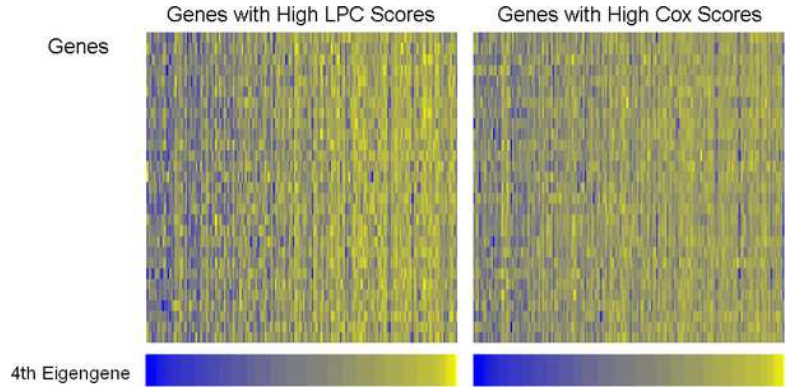

FIG. 11. *For the kidney data set, the top 29 genes selected by LPC but not by Cox score are shown on the top left, and the top 29 genes selected by Cox score but not by LPC are shown on the top right. On the bottom left and the bottom right, the fourth eigengene is shown. The observations are ordered from left to right by the size of their loading in the fourth eigengene. The top 29 genes selected by LPC resemble the fourth eigengene more closely than do the top 29 genes selected by Cox score.*

3.5. *Loss of power if assumptions are incorrect.* Throughout this paper we have motivated LPC by arguing that, for gene expression data, genes work together in pathways, and a clinical outcome is likely to be caused not by a single gene, but rather by a set a correlated genes. An obvious question arises: How much power to detect significant genes will we lose using LPC if the response is caused by a single or a few genes, rather than by a large pathway of genes? In this case, we might expect that univariate gene scores would correctly identify the significant gene, whereas LPC applied to those gene scores would identify the significant gene, along with other correlated but unimportant genes.

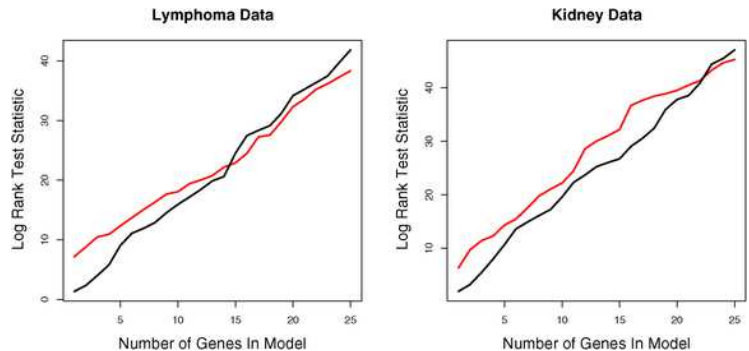

FIG. 12. *Cox proportional hazard models were fit to the test data, using the 1 through 25 genes that received highest Cox/LPC scores on the training data. When the number of genes in the model is low, the models fit using genes with high LPC scores had higher log rank statistics that those fit using genes with high Cox scores. LPC is red and T is black.*

As an example, we randomly selected 500 genes from the colon cancer data set, and performed $k$-means clustering on the genes with $k = 50$. We simulated a quantitative response $y$ as the centroid of the cluster containing the greatest number of genes, plus noise $(2N(0, 1))$. The resulting predictive advantage of LPC over $T$ was positive. We then repeated the experiment, but this time simulated the response as the centroid of the cluster containing the fewest genes, plus noise. The resulting predictive advantage of LPC over $T$ was negative. The predictive advantages for the two cases are displayed in Figure 1 in the Supplementary Materials [Witten and Tibshirani (2008)].

This example suggests that LPC has the potential to outperform $T$ in cases where the response is caused by a large pathway of correlated genes; however, if this condition is not satisfied, then $T$ may do better. As pointed out by an editor, it is likely that LPC will perform well in cases such as tumorigenesis that involve large perturbations to molecular pathways; on the other hand, LPC may not perform as well when more subtle changes in gene expression are present. If one does not know whether the use of LPC is warranted in a given situation, a simple plot of the predictive advantage will indicate whether LPC provides a benefit over $T$.

**4. Discussion.** Many of the ideas in this paper build upon the existing literature. Eigenarrays and eigengenes have been used many times before in the analysis of microarray data; for instance, Alter, Brown and Botstein (2000) decompose microarray data into eigengenes and eigenarrays in order to obtain a "global picture of the dynamics of gene expression," and Bair and Tibshirani (2004) and Bair et al. (2006) use eigengenes in order to predict patient survival time. Leek and Storey (2007) use the eigengenes of microarray data in order to infer and remove expression heterogeneity that is unrelated to the outcome. (More on the relationship between LPC and surrogate variable analysis can be found in Appendix D.2.) In addition, Shen et al. (2006) make use of the singular value decomposition for the selection of genes to use in tumor classification. Unlike LPC, Shen et al. (2006) seek genes that are uncorrelated with each other and capture the variation in the data, rather than genes that are associated with the outcome. To the best of our knowledge, we are the first to propose regressing simple gene scores onto the eigenarrays in order to obtain gene scores that make use of the data's correlation structure and association with the outcome.

The notion of using latent factors and sparse representations in order to model biological pathways is well developed in the statistical literature. In particular, West (2003) and Carvalho et al. (2008) use a Bayesian approach to fit sparse factor models to gene expression data. They demonstrate that these sparse factors are easily interpreted, and can lead to the discovery and validation of biological pathways. Their method can also be used for gene selection: if a few factors are associated with the outcome, then this suggests that only genes with nonzero loadings in those factors are related to the outcome. Our approach is different, in that we use "empirical factors" rather that attempting to model the factors themselves in a sparse way. It would be of interest to combine these two approaches by regressing univariate statistics onto these sparse factors, rather than onto the eigenarrays of the data as described here.

LPC takes advantage of the structure of the entire microarray data set in order to improve accuracy of gene scores. In order to achieve this same goal, one might instead consider the use of a full multivariate approach. For instance, one could regress a quantitative outcome onto the gene expression matrix using an $L_2$ (ridge) penalty. The resulting coefficient for each gene could be treated as a measure of its differential expression. However, suppose that some genes in the expression data matrix are highly correlated with each other and with the outcome. Ridge regression would assign to all of these genes similar coefficients that are smaller in magnitude than the coefficients that they each would receive in a univariate regression. In effect, ridge regression would decrease the estimate of a gene's significance as a consequence of the presence of correlated genes. Similarly, one could regress the outcome onto the data matrix with an $L_1$ penalty. Due to the sparse nature of the $L_1$ solution, this would result in only one gene out of a correlated set of genes receiving a nonzero coefficient. From these examples, it is clear that such multivariate approaches are not well suited to the problem of identifying differentially-expressed genes.

An attribute of LPC is that its use is not limited to the identification of significant genes in microarray data. Rather, it can be applied to a wide range of data types, provided that the hypothesis that significant features occur in correlated sets is justified. Interestingly, a recent paper by Price et al. (2006) makes use of a technique that is similar to LPC for a completely different reason. In the analysis of genome-wide association data, in order to identify single-nucleotide polymorphisms (SNPs) that are associated with a given phenotype, Price et al. (2006) suggest "adjusting" the phenotypes and genotypes by regressing them onto the principal components of the genotype data, and using the residuals to compute a $\chi^2$ statistic to measure the association between a given SNP and the phenotype. They do this to remove the effects of ancestry from the data. In effect, LPC seeks features that are associated with the principal components of the data, whereas Price et al. (2006) do the opposite. This is due to the two very different biological processes underlying the two types of data. In the case of microarray data, we seek groups of genes that are correlated with each other and with the outcome; in the case of genome-wide association data, we seek a single causative SNP or set of interacting SNPS that affect disease risk.

In practice, the LPC method can be used in two ways. The easiest approach, which is discussed throughout most of the paper, is to apply it to simple $T$ scores. As shown here, the resulting scores can result in lower FDR than much more complicated scores, such as ODP. Another possibility is that the LPC method can be applied directly to more complicated scores such as ODP. (This was done in Figure 4.) This can lead to even greater decreases in FDR than simply applying LPC to $T$, and is an attractive option in situations where there is no need to explain the gene scores to a nonstatistical audience. Though LPC outperforms $T$ on the simulated and real data examples that are shown here, it does not always provide an improvement over $T$. For microarray data, we recommend the use of LPC instead of a competing method whenever its predictive advantage, relative to the competing method, is positive.

An **R** software package will be made available.

## APPENDIX A: SELECTION OF THE SHRINKAGE PARAMETER

The LPC method involves a tuning parameter, $\lambda$. We choose the value of $\lambda$ via cross-validation, as follows:

**Selection of $\lambda$:**

1. Compute $\mathbf{V}$, the matrix of eigenarrays of $\mathbf{X}$.
2. Split the observations into a training set and a test set.
3. For a range of values of $\lambda$:
   (a) Compute $\mathbf{L}_{\lambda,\text{train}}$, which are similar to the LPC scores described in the LPC Algorithm, except that the matrix $\mathbf{V}$ on which the $\mathbf{T}_{\text{train}}$ scores are regressed is taken from Step 1 above. In other words, we regress the training set $T$ scores onto the eigenarrays for the entire data set.
   (b) Compute the average $|\mathbf{T}_{\text{test}}|$ of the 50 genes with the highest $|\mathbf{L}_{\lambda,\text{train}}|$ scores.
4. Repeat steps 2 and 3 ten times, recording the average $|\mathbf{T}_{\text{test}}|$ scores obtained in step 3(b).
5. Choose the value of $\lambda$ that results in the highest average $|\mathbf{T}_{\text{test}}|$ of the 50 genes with the highest $|\mathbf{L}_{\lambda,\text{train}}|$ scores.

We regress the training set $T$ scores onto the eigenarrays of the full data set, rather than onto the eigenarrays of the training data set, for two reasons. First of all, this leads to a much faster algorithm, as it means that we are required to compute the SVD only once, for the full data set, rather than once for each of the 10 training sets that we produce. In addition, if we were to regress the test set $T$ scores onto the eigenarrays for the training set, we would not be guaranteed that those eigenarrays resemble those of the full data set. Therefore, it would not be clear that the optimal parameter value for the split data sets is also best for the full data set. Note that we are not over-fitting the data by using the eigenarrays for the full data set, since computing the eigenarrays does not involve the test set $\mathbf{y}$ values.

Our method of choosing $\lambda$ is closely related to the concept of predictive advantage, presented in Section 3.2. It is worth noting that in most of our simulations, LPC's performance was not very sensitive to the choice of $\lambda$: for a wide range of $\lambda$ values, LPC outperformed $T$ by a comfortable margin.

## APPENDIX B: LPC FOR A MULTIPLE-CLASS RESPONSE

In the case of a multiple-class outcome with $K$ classes, we compute LPC scores in a slightly different fashion from the method described in the main text:

**LPC Algorithm for Multiple-Class Outcome:**

1. Compute the contrast $S_{kj}$ for each class $k$ (which we will call $C_k$) and for each gene $j$: $S_{kj} = \frac{\bar{\mathbf{X}}_{C_k j} - \bar{\mathbf{X}}_j}{s_j}$, where $s_j$ is the standard deviation for gene $j$. $\mathbf{S}_k$ denotes the vector of contrasts for class $k$.
2. Compute $\mathbf{v}_1, \ldots, \mathbf{v}_n$, the eigenarrays of $\mathbf{X}$, where the $\mathbf{v}_i$'s are the columns of the matrix $\mathbf{V}$ in the singular value decomposition (SVD) of $\mathbf{X}$, $\mathbf{X} = \mathbf{U}\mathbf{D}\mathbf{V}^T$.
3. For some value of the parameter $\lambda$, fit the model $\mathbf{S}_k = \boldsymbol{\beta}_0 + \sum_{i=1}^{n} \beta_i \mathbf{v}_i$, where the vector $\boldsymbol{\beta}$ with components $\beta_i$ is chosen to minimize the quantity $(\mathbf{S}_k - \boldsymbol{\beta}_0 - \mathbf{V}\beta)^T (\mathbf{S}_k - \boldsymbol{\beta}_0 - \mathbf{V}\boldsymbol{\beta}) + \lambda \sum_{i=1}^{n} |\beta_i|$.
4. Let $\hat{S}_{kj}$ denote the fitted values obtained by the above model for class $k$ and gene $j$. The LPC score for gene $j$ is $\sum_{k=1}^{K} \hat{S}_{kj}^2$.

Note that the two-class case really is a special case of the multiple-class case with $K = 2$, where the LPC score for gene $j$ is given by $\hat{S}_{1j}$.

## APPENDIX C: ESTIMATION OF FALSE DISCOVERY RATE FOR LPC

**C.1. The problem of functionally dependent statistics.** Estimation of the false discovery rate for the LPC procedure is surprisingly difficult. In studying this issue, we learned a more general fact about FDR estimation, which we discuss here. This issue is also addressed in Getz et al. (2007).

TABLE 2
*Possible outcomes from a multiple testing problem*

|      |         | Predicted |         |
|------|---------|-----------|---------|
|      |         | **Null**  | **Nonnull** |
| True | Null    | A         | B       |
|      | Nonnull | C         | D       |

False discovery rates are often estimated by permutations, which are simulated from a null distribution. This is done, for example, in the SAM procedure [Tusher, Tibshirani and Chu (2001)]. It turns out that the validity of this procedure relies on the fact that the statistic for gene $j$ is a function of only the data for gene $j$. This is violated for the LPC score, which is functionally dependent on all of the data for all of the genes.

Consider the usual testing scenario, with outcomes given in Table 2. The FDR is defined as $\mathrm{E}(B/(B+D))$. This expectation is taken with respect to the population of genes, which are null and nonnull; it is not taken with respect to a null distribution.

Now, null simulations try to estimate the FDR by simulating data from the top row of this table, counting the average number of false positives $B^*$, and then using the approximation $\widehat{FDR} = (\mathrm{ave}\,\#B^*)/(B+D)$. This assumes that $\mathrm{E}(B) \approx \mathrm{E}_{\mathrm{null}}(B)$ (the expectation under the null). This is true for statistics that are functions of just one gene, but is not true for functionally dependent statistics. In the latter case, the interplay between null and nonnull genes creates a large bias in $\mathrm{E}_{\mathrm{null}}(B)$ as an estimate of $\mathrm{E}(B)$.

Figure 2 in the Supplementary Materials [Witten and Tibshirani (2008)] shows a simple example with $p$-values. We generated 1000 $p$-values with a $U[0,1]$ distribution and then set the first 50 to $10^{-6}$. These are shown in the top left panel: the spike on the left are the nonnull $p$-values. We generated 1000 new $p$-values from $U[0,1]$, shown in the top right panel. As we expect, their distribution looks like that of the null $p$-values in the top left. For illustration, we chose a cutoff of $t_0 = 0.054$, the 10% point of the histogram in the top left panel. Hence, 100 $p$-values in the top left panel are less than 0.054. We simulated 200 versions of the top right panel, and the average number of $p$-values that were below $t_0$ was 54.4. Hence, the estimated FDR is $54.4/100 = 0.544$. The true FDR is 50%, since only one-half of these genes are nonnull. As we expect, the estimated FDR is close to the actual FDR.

As an illustration, suppose that we instead use a simple transformation of the $p$-value,

$$\text{(C.1)} \qquad \mathrm{CaMP}_j = -\log_{10}(np_j/q_j),$$

where $q_j$ is the rank of the $p_j$ among all of the $p$-values. This score was used in Sjoblom et al. (2006): it has some appeal from an interpretability viewpoint, but also creates a functional dependence between the $p$-values. The bottom two panels of the figure show what happens when we repeat the exercise with $CaMP_j$ instead of $p_j$. (Note that large values of $CaMP_j$ indicate significance.) The histogram in the bottom right is shifted to the left compared to that for the null CaMP scores in the bottom left panel. Using a 90% cutoff of 0.236 for the CaMP score, the estimated FDR is 0.017, far less than the actual FDR of 50%.

What has happened? The CaMP scores for the null genes behave differently depending on whether or not there are nonnull scores in the data, because the scores are functionally dependent. The presence of nonnull scores in the bottom left panel tends to inflate the scores for the null genes.

This same phenomenon occurs if we use permutations to estimate the null distribution for an arbitrary statistic that is a function of multiple genes. If the permutation approach is used, the bias in the estimate of the FDR can result in over-estimation or under-estimation of the true FDR. The LPC statistic is a functionally dependent statistic, since the data for all genes is used to estimate the principal components. Therefore, the permutation approach is not suitable for the estimation of the FDR for LPC.

**C.2. Estimating FDR via predictive advantage.** The FDR of $T$ can be easily estimated via permutations. Therefore, it makes sense to try to assess the FDR of the LPC statistic relative to the FDR of $T$. We take that approach in this section.

It makes sense intuitively that an estimator with higher predictive advantage (3.1) will also tend to have lower FDR. In this section we show that under a simple shift model, the FDR of a statistic is lower than that of $T$ if the statistic has higher predictive advantage. Second, we derive an estimate of the FDR of LPC based on an adjustment of the FDR of $T$. These results hold for any functionally dependent statistic, not just the LPC score.

We use the same notation as in previous sections. Let $T_j$ be the $T$ score for gene $j$ on the training set, and $L_j$ the LPC score for gene $j$ on the training set. On the test set, the $T$ score for gene $j$ is denoted $T_j^*$. Let $c_1(\alpha)$ and $c_2(\alpha)$ denote the $\alpha$ quantiles of $T_j$ and $L_j$. Also, let $T_j = u_j + z_j$ and $T_j^* = u_j + z_j^*$, where $u_j$, $z_j$ and $z_j^*$ are independent, $u_j = 0$ if gene $j$ is null, and $z_j$ and $z_j^*$ are identically distributed.

We wish to show that

(C.2) $$P(T_j^* > c | L_j > c_2(\alpha)) > P(T_j^* > c | T_j > c_1(\alpha))$$

implies

(C.3) $$P(u_j = 0 | L_j > c_2(\alpha)) < P(u_j = 0 | T_j > c_1(\alpha)).$$

Now,

$$P(T_j^* > c | L_j > c_2(\alpha))$$
$$= P(T_j^* > c | L_j > c_2(\alpha), u_j = 0) P(u_j = 0 | L_j > c_2(\alpha))$$
(C.4)
$$\quad + P(T_j^* > c | L_j > c_2(\alpha), u_j \neq 0) P(u_j \neq 0 | L_j > c_2(\alpha))$$
$$= P(T_j^* > c | L_j > c_2(\alpha), u_j = 0) P(u_j = 0 | L_j > c_2(\alpha))$$
$$\quad + P(T_j^* > c | L_j > c_2(\alpha), u_j \neq 0)[1 - P(u_j = 0 | L_j > c_2(\alpha))].$$

So,

(C.5)
$$P(u_j = 0 | L_j > c_2(\alpha))$$
$$= \frac{P(T_j^* > c | L_j > c_2(\alpha), u_j \neq 0) - P(T_j^* > c | L_j > c_2(\alpha))}{P(T_j^* > c | L_j > c_2(\alpha), u_j \neq 0) - P(T_j^* > c | L_j > c_2(\alpha), u_j = 0)}.$$

Now, we make the additional assumption that $u_j = 1$ if a gene is nonnull (recall that we have already assumed that $u_j = 0$ if a gene is null). It follows that

(C.6)
$$P(u_j = 0 | L_j > c_2(\alpha)) = \frac{P(T_j^* > c | u_j = 1) - P(T_j^* > c | L_j > c_2(\alpha))}{P(T_j^* > c | u_j = 1) - P(T_j^* > c | u_j = 0)}.$$

Similarly, we can expand $P(T_j^* > c | T_j > c_1(\alpha))$ to find that

(C.7)
$$P(u_j = 0 | T_j > c_1(\alpha)) = \frac{P(T_j^* > c | u_j = 1) - P(T_j^* > c | T_j > c_1(\alpha))}{P(T_j^* > c | u_j = 1) - P(T_j^* > c | u_j = 0)}.$$

Note that (C.6) and (C.7) have the same denominator, which is positive. And, from (C.2), we know that the numerator of (C.6) is less than the numerator of (C.7) (and the numerators are positive). So, (C.7) is greater than (C.6); in other words,

(C.8)
$$P(u_j = 0 | T_j > c_1(\alpha)) > P(u_j = 0 | L_j > c_2(\alpha)).$$

So, the FDR of LPC is bounded above by the FDR of $T$. This result assumes that there is only one parameter value in the alternative space, and does not exactly hold in the composite case.

To estimate the FDR, we use (C.6) and (C.7). We approximate the difference in FDRs by

(C.9)
$$P(u_j = 0 | T_j > c_1(\alpha)) - P(u_j = 0 | L_j > c_2(\alpha))$$
$$= (1 - \pi_0) \cdot \frac{P(T_j^* > c | L_j > c_2(\alpha)) - P(T_j^* > c | T_j > c_1(\alpha))}{P(T_j^* > c) - P(T_j^* > c | u_j = 0)},$$

where $\pi_0$ is the proportion of genes in the data set that are not significant.

We deal with $c$ in the computations by using the relation $\mathrm{E}(X) = \int_0^\infty (1 - F(x))\,dx$ for a positive random variable. Hence, we average the quantities on the right-hand side (e.g., $T_j^*$).

Figure 3 in the Supplementary Materials [Witten and Tibshirani (2008)] shows the estimated FDRs for LPC and $T$ for Simulations 1, 2, and 3 with a quantitative response variable. The estimate of FDR for LPC [using (C.9)] is pretty accurate. We have found that it behaves fairly well in general, and it tends to err in the conservative direction.

## APPENDIX D: AN UNDERLYING LATENT VARIABLE MODEL

**D.1. Model and results.** Assume that the data are generated under the following model, in which the response $\mathbf{y}$ is quantitative:

(D.1)
$$y_i = \beta_0 + \beta_1 u_{i1} + \varepsilon_i,$$
$$X_{ij} = \alpha_{0j} + c_1 \alpha_{1j} u_{i1} + c_2 \alpha_{2j} u_{i2} + z_{ij},$$
$$\mathrm{E}(\varepsilon_i) = \mathrm{E}(z_{ij}) = \mathrm{E}(\varepsilon_i z_{ij}) = 0,$$

where $P$ denotes the set of important genes, and $\alpha_{1j} = 0$ if $j \notin P$, $\alpha_{1j} \neq 0$ if $j \in P$. Let $\mathbf{u}_1 = (u_{11} \cdots u_{n1})^T$ and $\mathbf{u}_2 = (u_{12} \cdots u_{n2})^T$ be orthonormal with mean zero. Similarly, let $\boldsymbol{\alpha}_1 = (\alpha_{11} \cdots \alpha_{1p})^T$ and $\boldsymbol{\alpha}_2 = (\alpha_{21} \cdots \alpha_{2p})^T$ be orthonormal with mean zero. Assume that $\mathbf{y}$ has been centered to have mean zero so that $\beta_0 = 0$, and that the genes are centered, so that $\alpha_{0j} = 0$. Also, assume that the $z_{ij}$ are independent and identically distributed for all $i \in 1, \ldots, n$ and $j \in 1, \ldots, p$, and that the $\varepsilon_i$ are independent and identically distributed for all $i \in 1, \ldots, n$.

First, we consider simplified $T$ scores that take the form $T_j = \mathbf{X}_j^T \mathbf{y}$ for gene $j$ (these simplified scores are equivalent to the usual $T$ scores one would obtain if one first scaled the columns of $\mathbf{X}$ appropriately). Then,

(D.2)
$$\mathrm{E}(T_j) = \mathrm{E}(\mathbf{X}_j^T \mathbf{y})$$
$$= \sum_{i=1}^n \mathrm{E}((c_1 \alpha_{1j} u_{i1} + c_2 \alpha_{2j} u_{i2} + z_{ij})(\beta_1 u_{i1} + \varepsilon_i))$$
$$= c_1 \alpha_{1j} \beta_1,$$

because $\mathrm{E}(z_{ij}) = \mathrm{E}(\varepsilon_i) = \mathrm{E}(z_{ij}\varepsilon_i) = 0$, and $\mathbf{u}_1$ and $\mathbf{u}_2$ are orthonormal with mean zero.

Now, suppose that we ignore the error associated with estimation of the principal components, so that we estimate some eigenarray $\mathbf{v}$ of the data matrix $\mathbf{X}$ to equal $\boldsymbol{\alpha}_1 = (\alpha_{11} \cdots \alpha_{1p})^T$ from the underlying model. We can regress $\mathbf{X}^T \mathbf{y}$ onto $\mathbf{v} = (v_1 \cdots v_p)$ in order to obtain LPC scores:

(D.3)
$$\hat{\mathbf{T}} = \langle \mathbf{X}^T \mathbf{y}, \mathbf{v} \rangle \mathbf{v},$$

where $\langle \cdot , \cdot \rangle$ denotes inner product, and $\widehat{\mathbf{T}} = (\widehat{T}_1 \cdots \widehat{T}_p)^T$. We want the expectation of $\widehat{T}_j$:

$$\text{(D.4)} \qquad \text{E}(\widehat{T}_j) = \text{E}(v_j \langle \mathbf{X}^T \mathbf{y}, \mathbf{v} \rangle) = \alpha_{1j} c_1 \beta_1 \alpha_1^T \boldsymbol{\alpha}_1 = \alpha_{1j} c_1 \beta_1 = \text{E}(T_j).$$

Now, to find the variance of $\widehat{T}_j$,

$$
\begin{aligned}
\text{Var}(\widehat{T}_j) &= \text{Var}(v_j \langle \mathbf{X}^T \mathbf{y}, \mathbf{v} \rangle) \\
&= \alpha_{1j}^2 \, \text{Var}(\boldsymbol{\alpha}_1^T \mathbf{X}^T \mathbf{y}) \\
&= \alpha_{1j}^2 \alpha_1^T \, \text{Var}(\mathbf{X}^T \mathbf{y}) \boldsymbol{\alpha}_1 \\
&= \alpha_{1j}^2 \alpha_1^T \, \text{Var}(\mathbf{T}) \boldsymbol{\alpha}_1.
\end{aligned}
$$

(D.5)

Now,

$$\text{(D.6)} \quad \text{Var}(\mathbf{T}) = (c_1^2 \alpha_1 \alpha_1^T + c_2^2 \alpha_2 \alpha_2^T) \, \text{Var}(\varepsilon_i) + I_{p \times p}(\beta_1^2 \, \text{Var}(z_{ij}) + \text{Var}(\mathbf{z}_j^T \boldsymbol{\varepsilon}))$$

and so

$$
\begin{aligned}
\text{Var}(\widehat{T}_j) &= c_1^2 \alpha_{1j}^2 \, \text{Var}(\varepsilon_i) + \alpha_{1j}^2 \beta_1^2 \, \text{Var}(z_{ij}) + \alpha_{1j}^2 \, \text{Var}(\mathbf{z}_j^T \boldsymbol{\varepsilon}), \\
\text{Var}(T_j) &= (c_1^2 \alpha_{1j}^2 + c_2^2 \alpha_{2j}^2) \, \text{Var}(\varepsilon_i) + \beta_1^2 \, \text{Var}(z_{ij}) + \text{Var}(\mathbf{z}_j^T \boldsymbol{\varepsilon}), \\
\text{Var}(T_j) - \text{Var}(\widehat{T}_j) &= c_2^2 \alpha_{2j}^2 \, \text{Var}(\varepsilon_i) + (1 - \alpha_{1j}^2)[\beta_1^2 \, \text{Var}(z_{ij}) + \text{Var}(\mathbf{z}_j^T \boldsymbol{\varepsilon})].
\end{aligned}
$$

(D.7)

Therefore, using $\widehat{T}_j$ rather than $\mathbf{X}_j^T \mathbf{y}$ to rank the significant genes results in a decrease in variance (note that we have assumed that $\mathbf{v} = \boldsymbol{\alpha}_1$; we are ignoring the variance associated with the estimation of the eigenarrays). Equation (D.4) indicates that the LPC scores and the $T$ scores have the same expectation.

In this example, only one latent factor ($\mathbf{u}_1$) is related to the response. As pointed out by a reviewer, the argument does not extend exactly to the case of multiple latent factors.

Details of the calculations performed in this section can be found in the Supplementary Materials [Witten and Tibshirani (2008)].

**D.2. Relation to surrogate variable analysis.** In a recent paper Leek and Storey (2007) present surrogate variable analysis (SVA), a novel method that incorporates sources of expression heterogeneity into microarray data analysis in order to increase power and mitigate spurious signals in genes that are unrelated to outcome. An editor noted that there are similarities between SVA and LPC. While both methods involve using the singular value decomposition in order to identify structure in the data shared by the response and the gene expression measurements, they have different effects.

Consider, as an example, the result of applying SVA to the data generated under our latent variable model (Appendix D.1). We use the same notation as in that section. SVA first detects "unmodeled factors" in the gene expression data that are not explained by the model

$$(D.8) \qquad X_{ij} = \mu_j + f_j(y_i) + \varepsilon_{ij}.$$

Let us assume that SVA is able to correctly identify the $\mathbf{u}_2$ term as a source of expression heterogeneity unrelated to the outcome. Then, in subsequent analysis, Leek and Storey (2007) suggest the use of the modified data

$$(D.9) \qquad \mathbf{X}_j^* = \mathbf{X}_j - c_2 \alpha_{2j} \mathbf{u}_2,$$

which is what remains after removing the surrogate variable that we estimated. The conventional scores using the modified data (call them $T^*$) will take the form

$$
\begin{aligned}
(D.10) \qquad T_j^* &= \mathbf{X}_j^{*T} \mathbf{y} \\
&= (\mathbf{X}_j - c_2 \alpha_{2j} \mathbf{u}_2)^T (\beta_1 \mathbf{u}_1 + \boldsymbol{\varepsilon}) \\
&= (c_1 \alpha_{1j} \mathbf{u}_1 + \mathbf{z}_j)^T (\beta_1 \mathbf{u}_1 + \boldsymbol{\varepsilon}).
\end{aligned}
$$

It is clear that $\mathrm{E}(T_j^*) = \mathrm{E}(T_j)$. As a consequence of our assumption that we were able to successfully identify the surrogate variable $\mathbf{u}_2$, it follows that $\mathrm{Var}(T_j^*) = c_1^2 \alpha_{1j}^2 \mathrm{Var}(\varepsilon_i) + \beta_1^2 \mathrm{Var}(z_{ij}) + \mathrm{Var}(\mathbf{z}_j^T \boldsymbol{\varepsilon}) < \mathrm{Var}(T_j)$. So, under our model, SVA has successfully reduced the variance of the $T$ scores.

However, at this point, the SVA scores $T^*$ can be regressed onto the eigenarrays of the original data, as in the LPC method, in order to obtain a reduction in variance as in Section D.1. In fact, by the argument in the previous section, the variance of LPC applied to $T^*$ is the same as that of LPC applied to $T$ (and is less than the variance of $T^*$). Therefore, in this example, LPC provides benefits in variance reduction that are beyond those given by SVA. In general, the LPC method can be applied on top of gene scores obtained after expression heterogeneity has been removed using SVA.

## APPENDIX E: SIMULATION DETAILS

**E.1. Description of simulations.** Here, we describe in greater details the simulations used in Section 3.1.

Our first simulation represents the simplest case: all of the signal in the data is associated with the outcome.

**Simulation 1: Simple Case**

1. In observations 1–20, $y_i \sim N(6, 1)$, and in observations 21–40, $y_i \sim N(5, 1)$.
2. In observations 1–20, $X_{ij} \sim N(2, 1)$ for $j \le 50$. Otherwise (for $i > 20$ and for $j > 50$), $X_{ij} \sim N(0, 1)$.

In our second simulation the significant genes have less signal than large blocks of genes that are unrelated to the outcome.

**Simulation 2: Blocks of Noise Genes**

1. In observations 1–20, $y_i \sim N(12.5, 1)$, and in observations 21–40, $y_i \sim N(10, 1)$.
2. In observations 1–20 and in genes 1–50, $X_{ij} \sim N(1.5, 1)$.
3. For all other $i, j$, we let $X_{ij} \sim N(0, 1)$, with the following exception: there are three blocks of 100 genes each that are distributed $N(2, 1)$ or $N(-2, 1)$ in 10 observations; these three blocks of genes are orthogonal to the signal present in the outcome and in the first 50 genes.

The blocks of noise genes represent groups of genes with strong expression patterns that are unrelated to the outcome of interest.

In our final simulation there are two sets of 25 significant genes that have orthogonal expression patterns. The quantitative outcome is the sum of the expression patterns in the two sets of significant genes.

**Simulation 3: Two Sets of Orthogonal Significant Genes**

1. For observations 1–10, $y_i \sim N(10, 1)$; for observations 11–30, $y_i \sim N(11, 1)$, and for observations 31–40, $y_i \sim N(12, 1)$.
2. For genes 1–25, $X_{ij} \sim N(0, 1)$ for observations 1–20 and $X_{ij} \sim N(2, 1)$ for observations 21–40.
3. For genes 26–50, $X_{ij} \sim N(0, 1)$ for observations 1–10 and 21–30. For observations 11–20 and 31–40, $X_{ij} \sim N(2, 1)$.
4. For genes 51–1000, $X_{ij} \sim N(0, 1)$.

Heatmaps for each simulation can be seen in Figure 4 in the Supplementary Materials [Witten and Tibshirani (2008)]. The simulations described above have a quantitative outcome; two-class outcomes are obtained using an indicator variable for whether the outcome for a given observation is greater than or less than the median outcome across all observations.

**Acknowledgments.** We would like to thank Holger Hoefling for valuable discussions of the false discovery rate issues, and Trevor Hastie, Gen Nowak, two referees and editors for helpful comments.

## SUPPLEMENTARY MATERIAL

DEPARTMENT OF STATISTICS
STANFORD UNIVERSITY
390 SERRA MALL
STANFORD, CALIFORNIA 94305
USA
E-MAIL: dwitten@stanford.edu

DEPARTMENTS OF HEALTH RESEARCH
AND POLICY AND STATISTICS
STANFORD UNIVERSITY
390 SERRA MALL
STANFORD, CALIFORNIA 94305
USA
E-MAIL: tibs@stat.stanford.edu